\documentstyle[aaspp4]{article}
\begin{document}

\newcommand\mass{\ifmmode{\cal M}\else{${\cal M}$}\fi}
\newcommand\OL{\ifmmode{\Omega_{\Lambda}}\else{$\Omega_{\Lambda}$}\fi}
\newcommand\Ob{\ifmmode{\Omega_{b}}\else{$\Omega_{b}$}\fi}
\newcommand\Om{\ifmmode{\Omega_{m}}\else{$\Omega_{m}$}\fi}
\newcommand\Oc{\ifmmode{\Omega_{CDM}}\else{$\Omega_{CDM}$}\fi}
\newcommand\On{\ifmmode{\Omega_{\nu}}\else{$\Omega_{\nu}$}\fi}
\newcommand\LCDM{$\Lambda$CDM}
\newcommand\boom{BOOMERanG}
\newcommand\prior{{\it a priori\/}}

\title{BOOMERanG Data Suggest a Purely Baryonic Universe}
\author{Stacy S. McGaugh}
\affil{Department of Astronomy, University of Maryland,
        College Park, MD 20742-2421 \\
        e-mail: ssm@astro.umd.edu}

\begin{abstract}
The amplitudes of peaks in the angular power
spectrum of anisotropies in the microwave background radiation
depend on the mass content of the universe.  
The second peak should be prominent when cold dark matter is
dominant, but is depressed when baryons dominate.  
Recent microwave background data are consistent with a
purely baryonic universe with $\Om = \Ob$ and $\OL \approx 1$. 
\end{abstract}

\section{Introduction}

At present, the standard cosmological paradigm is a universe in which
ordinary matter is a minor constituent, with $\sim 90\%$ of the mass
being in some non-baryonic form.  This is usually presumed to be some
new fundamental particle (e.g., WIMPs or axions), which in the astronomical
context is generically referred to as cold dark matter (CDM).  ``Standard''
CDM began as a compelling and straightforward
theory with few moving parts (e.g., Blumenthal et al.\ 1984\markcite{BFPR}).
It has evolved into a model (\LCDM) with many fine tuned
parameters (e.g., Ostriker \& Steinhardt 1995\markcite{OS}).  This might
reflect our growing knowledge of real complexities, or it might be a sign
of some fundamental problem.

As yet, we have no direct indication that CDM actually exists. 
Consequently, the assumption that it makes up the vast majority of mass
in the universe remains just that:  an assumption.  The presumed existence of
CDM is a well motivated inference based principally on two astrophysical
observations.  One is that the total mass density inferred dynamically
greatly exceeds that allowed for normal baryonic matter by big bang
nucleosynthesis ($\Omega_m > \Ob$).  The other is that the cosmic microwave
background is very smooth.  Structure can not grow gravitationally to the
rich extent seen today unless there is a non-baryonic component which can
already be significantly clumped at the time of recombination without leaving
incriminatingly large fingerprints on the microwave background.

Nevertheless, CDM faces some severe problems, especially at smaller scales
(e.g., Moore 1994\markcite{Ben}; Flores \& Primack 1994\markcite{FP};
McGaugh \& de Blok 1998a\markcite{MBa}; Moore et al.\ 1999\markcite{MQGSL};
Navarro \& Steinmetz 2000\markcite{NS}; Sellwood 2000\markcite{Jerry}).  
Since the existence of CDM remains an assumption,
it seems prudent to consider the case of a purely baryonic universe. 
In this context, it is not surprising that the second peak is
constrained to have a small amplitude in the data reported by recent
microwave background experiments (de Bernardis et al.\ 2000\markcite{Boom};
Hanany et al.\ 2000\markcite{maxima}).  
It is expected (McGaugh 1999\markcite{mypred}).

\section{Prior Predictions}

Models for the angular power spectrum of fluctuations in the microwave
background have many free parameters (Seljak \& Zaldarriaga\markcite{cmbfast}
1996).  Many of these parameters are degenerate
(Efstathiou \& Bond 1999\markcite{EB}), making it possible to fit
a wide variety of models to any given data set (e.g.,
Lange et al.\ 2000\markcite{lange}).  This makes the
role of prior constraints, and \prior\ predictions, particularly important.

Fortunately, the baryon content is the principal component which affects
the relative amplitude of the even and odd peaks.  For the baryon content
specified by the abundances of the light elements and big bang nucleosynthesis
(e.g., Tytler et al.\ 2000\markcite{tyt}), both should be present.  However,
the even numbered rarefaction peaks should be more prominent when
CDM dominates the mass budget.  When it does not, baryonic drag
suppresses their amplitude
(Hu, Sugiyama, \& Silk\markcite{HSS} 1997).  As \Oc\ declines, the
amplitude of the second peak declines with it.
In the case where $\Oc \rightarrow 0$, the second peak is expected to
have a much smaller amplitude than in \LCDM\
(McGaugh 1999\markcite{mypred}),
consistent with the hints of a small secondary peak in
the \boom\ (de Bernardis et al.\ 2000\markcite{Boom}) and MAXIMA-1 data
(Hanany et al.\ 2000\markcite{maxima}).
  
The \prior\ predictions for the standard \LCDM\ paradigm and the pure baryon
case (McGaugh 1999\markcite{mypred}) are shown together with the \boom\
data\footnote{There is a significant zero-point offset between 
\boom\ and MAXIMA-1.  To rectify this, one must choose an arbitrary scaling
factor (Hanany et al.\ 2000\markcite{maxima}).
I have therefore refrained from combining the two data sets.
It is the shape of the power spectrum, and not its normalization, which
is important here.  The two data sets are consistent in this respect.}
in Fig.\ 1.  In addition to the illustrative cases I published previously,
I have now carefully
chosen parameters (Table 1) which satisfy all the constraints which went
into building \LCDM\ in the first place (Ostriker \& Steinhardt\markcite{OS};
1995; Turner\markcite{debate} 1999), updated to include the recent estimate
of $\Ob h^2 = 0.019$ (Tytler et al.\ 2000\markcite{tyt}).
All reasonable variation of the parameters
which were considered in \LCDM\ prior to the \boom\ results 
significantly overpredict the amplitude of the second peak.
This is difficult to avoid as long as one remains consistent with
big bang nucleosynthesis and cluster baryon fractions
(Evrard 1997\markcite{Gus}; Bludman 1998\markcite{Sid}). 

In contrast, the \prior\ prediction for a purely baryonic universe is
consistent with the data (Fig.\ 1).  The amplitude of the second
peak\footnote{In McGaugh (1999\markcite{mypred}) I described the baryonic
models as having the second peak completely suppressed, with the third
peak appearing to be the second.  This is not correct.  Such a situation can
occur, but only for baryon-to-photon ratios greater than allowed by big bang
nucleosynthesis.  The second peak discussed there and here is indeed the
second (rarefaction) peak.  The difference between \LCDM\ and purely
baryonic models is in the amplitude of this peak.}
is predicted to be much lower than in universes dominated
by CDM, as observed.  The power spectra models in Fig.\ 1(b) are
identical to the models I published previously
(McGaugh 1999\markcite{mypred}).  The only difference is that I have
scaled the geometry to match the precise position of the first peak.
This mapping is effectively an adjustment of the angular
scale by a factor $\alpha$ so that $\ell \rightarrow \alpha \ell$ (Table 1).
The \boom\ data prefer a geometry which is marginally closed,
which leads to $\alpha < 1$.  This is equivalent to a small adjustment in the
value of \OL\ (Table 1).
Once the geometry is fixed, the rest follows.  It is in the shape of the power
spectrum, and not in the geometry, in which there is a test of the
presence or absence of CDM.  I have not adjusted the shape {\it at all\/} from
what I predicted in McGaugh (1999\markcite{mypred}): this is as close to
a ``no-hands'' model as one can come.  The pure baryon models provide
a good description of the data.

In addition to the models of McGaugh (1999\markcite{mypred}), I illustrate
in Fig.\ 1(b) a model which adheres to the most recent estimate of $\Ob h^2$
(Tytler et al.\ 2000\markcite{tyt}).  In this case I have adjusted \OL\
to match the position of the first peak so that $\alpha = 1$ (Table 1).
The shape of the power spectrum measured by the \boom\ experiment is well
predicted by taking strong priors for \Ob, $H_0$, and so on, with the most
important being the pure baryon prior $\Oc = 0$.  Simply scaling the 
pre-existing models with two fit parameters, the amplitude $\Delta T$
and the geometry, provides a good fit: $\chi_{\nu}^2 < 1$ (Table 1).  
The data are consistent with a cosmology in which\footnote{A small neutrino
mass $m_{\nu} \lesssim 1$ eV is also admissible.} $\Omega_m = \Ob$ and
$\OL \approx 1$.

\section{Quantitative Measures}

In order to make a fit-independent, quantitative prediction of
the differences expected between the \LCDM\ and pure baryon cases,
I proposed (McGaugh 1999\markcite{mypred}) several geometry independent
measures.  These are the ratio of positions of observed peaks 
$\ell_{n+1}/\ell_n$, the 
absolute amplitude ratio of the peaks $(C_{\ell,n}/C_{\ell,n+1})_{abs}$,
and the peak-to-trough amplitude ratio $(C_{\ell,n}/C_{\ell,n+1})_{rel}$.

Of these measures, the first is the least sensitive and the last is
the most sensitive.  The ratio of the positions of the first two peaks
is expected to differ by only a small amount.  Until this quantity is
accurately measured, it does not provide a strong test.  Should a second
peak appear in future data, it does not necessarily favor \LCDM\ --- a
second peak is expected in either case, in roughly the same position.
What does provide a clear distinction is the last measure, the peak-to-trough
amplitude ratio of the first two peaks.  This
distinguishes between a second peak which
stands well above the first trough, as expected with CDM, and one which does
not, as expected without it.

These measures are readily extracted from the \boom\ data.  They are
reported in Table 2, together with the \prior\ predictions of the \LCDM\ and
pure baryon cases.  The data clearly fall in the regime favored by the
pure baryon case.  

The result remains in the regime favored by the pure baryon case even if
we adjust strategically chosen pairs of
data points in the direction favorable to CDM.  For example, increasing
the amplitude of the point at $\ell = 500$ where the second peak should occur
in \LCDM\ by $1\sigma$ and decreasing by $1\sigma$ the amplitude
of the point at $\ell = 400$ where the trough should occur
does not suffice to move the result away from the range favored by the
pure baryon case.  This is more than
a $2\sigma$ operation, as it is a coordinated move which would also impact
surrounding data points.  The \boom\ data clearly favor the case of zero CDM.

\section{Other Solutions}

Shortly after the \boom\ results were announced, various papers appeared
which attempted to explain the observed lack of a second peak.  These
take advantage of the many free parameters which are available in models
of the microwave background.  One solution is to increase the baryon content 
rather than reduce the CDM content.  In order to retain CDM one significantly
violates either big bang nucleosynthesis constraints
(Tegmark \& Zaldarriaga 2000\markcite{TZ}) or cluster baryon fractions,
or both.  These were critical pieces of evidence which led to \LCDM;
it is not a trivial matter to dispose of them in order to force the new
data into compliance with the model {\it du jour\/}.

Another solution is to somehow erase the peaks subsequent to the first.
This can happen if the microwave background photons encounter a significant
optical depth, which requires substantial reionization at quite early times
(Miller 2000\markcite{Cole}; Peebles, Seager, \& Hu 2000\markcite{PSH}).
How this could come about is unclear.
There may also be decoherence of the ideal signal
(White, Scott, \& Pierpaol 2000\markcite{WSP}), in which case the microwave
background will retain little information of interest beyond the position
of the first peak.

These effects were not expected, and
it is not necessary to invoke any of them if CDM does not exist.
The small observed amplitude of the second peak is natural and
expected.  Nevertheless, any of these effects could occur.  
The physics is the same in either
case --- the only difference is the presence or absence of CDM.  It is
much easier to explain the low observed amplitude of the second peak without
CDM.  That does not mean optical depth or decoherence or some other mundane
effect need not matter in the purely baryonic case.

At present, a simple universe devoid of CDM suffices to explain the \boom\ data.
{\it If\/} the universe remains simple, the pure baryon case continues to make
clear predictions.  As data accumulate, the second peak should become clear.
It is only marginally suggested by the data so far, but it should resolve into
the shape predicted by the models in Fig.\ 1(b).  The amplitude of this second
peak will be smaller than the \prior\ expectations of \LCDM\ models.  Beyond
this, the power spectrum should continue to roll off to smaller angular scales
so that the third peak has a lower absolute amplitude than the second.

\section{Just Baryons}

The angular power spectrum of the recent microwave background data
favor a purely baryonic universe over one dominated by CDM.
Yet a conventional baryonic universe with $\Om = \Ob$ faces the same
problems mentioned in the introduction which led to the invention of CDM.
For one, $\Om > \Ob$: dynamical measures give a total mass density an order
of magnitude in excess of the nucleosynthesis constraint on the baryon
density.  The other is that the gravitational growth of structure is slow:
$\delta \sim t^{2/3}$.  This makes it impossible to grow large scale
structure from the smooth initial state indicated by the microwave background
within the age of the universe. 

These arguments are compelling, but are themselves based on the assumption
that gravity behaves in a purely Newtonian fashion on all scales.
A modification to the conventional force law might also suffice.
One possibility which is empirically motivated is the
modified Newtonian dynamics (MOND) hypothesized by Milgrom\markcite{M83}
(1983).  MOND supposes that for accelerations $a \ll a_0 \approx
1.2 \times 10^{-10}\;{\rm m}\,{\rm s}^{-2}$, the effective acceleration
becomes $a \rightarrow \sqrt{g_N a_0}$, where $g_N$ is the usual Newtonian
acceleration which applies when $a \gg a_0$.  There is no dark matter in
this hypothesis, so the observed motions must relate directly to the
distribution of baryonic mass through the modified force law.

MOND has had considerable success in predicting the dynamics of a
remarkably wide variety of objects.  These include
spiral galaxies (Begeman, Broeils, \& Sanders\markcite{BBS}
1991; Sanders\markcite{S96} 1996; Sanders \& Verheijen\markcite{SV} 1998),
low surface brightness galaxies (McGaugh \& de Blok\markcite{MBb} 1998b;
de Blok \& McGaugh\markcite{BM} 1998; McGaugh\markcite{BTF} et al.\ 2000),
dwarf Spheroidals (Milgrom\markcite{7dw} 1997; Mateo\markcite{mario} 1998),
giant Ellipticals (Sanders\markcite{S00} 2000), groups
(Milgrom\markcite{M98} 1998) and clusters of galaxies
(Sanders 1994\markcite{cl1},1999\markcite{cl2}), and large scale
filaments (Milgrom\markcite{M97} 1997). 
The empirical evidence which supports MOND is rather
stronger than is widely appreciated.

Moreover, MOND does a good job of explaining the two observations that
motivated CDM.  The dynamical mass is overestimated when purely Newtonian
dynamics is employed in the MOND regime, so rather than $\Omega_m > \Ob$
one infers $\Omega_m \approx \Ob$ (Sanders 1998\markcite{S98};
McGaugh \& de Blok\markcite{MBb} 1998b).  The early universe is dense,
so accelerations are high and MOND effects\footnote{Assuming $a_0$ is
constant.} do not appear until after recombination.  When they do,
structure grows more rapidly than with conventional gravity
(Sanders 1998\markcite{S98}), so the problem in going from a smooth
microwave background to a rich amount of large scale structure is also
alleviated.  Since everything is normal in the high acceleration regime,
all the usual early universe results are retained. 

In order to get the position of the first peak right, we must invoke the
cosmological constant in either the conventional or MOND case.  In the
former case, it was once hoped that there would be enough CDM that $\Om = 1$.
In the latter case, $\Lambda$ may have its usual meaning, or it
may simply be a place holder for whatever the geometry really is.  
One possible physical basis for MOND may be the origin of inertial mass in
the interaction of particles with vacuum fields.  A non zero cosmological
constant modifies the vacuum and hence may modify inertia
(Milgrom\markcite{M99} 1999).  In this context, it is interesting to note that
for the parameters indicated by the data, $\Om = \Ob$ and $\OL \approx 1$,
the transition from matter domination to $\Lambda$-domination is roughly
coincident with the transition to MOND domination.

The value of \OL\ indicated by this scenario is in marginal conflict with
estimates from high redshift supernovae (Riess et al.\ 1998\markcite{hiz};
Perlmutter et al.\ 1999\markcite{SNCP}). 
Modest systematic effects might be present in Type Ia
supernovae data which could reconcile these results.
It is difficult to tell at this early stage how significant the difference
between $\OL \approx 0.7$ and $\OL \approx 1$ really is.
Even if this difference is real, it may simply indicate the extent
to which MOND affects the geometry.  This is analogous to the 
variable-$\Lambda$ scenarios called Quintessence which have recently
been considered (e.g., Caldwell, Dave, \& Steinhardt, 1998\markcite{CDS}).

\section{Conclusions}

Prior to the publication of the data from recent microwave background
experiments, I had investigated the power spectrum of anisotropies
which would be expected for a purely baryonic
universe devoid of CDM (McGaugh 1999\markcite{mypred}).  
Such a cosmology predicts a small amplitude for the second peak.
This prediction is consistent with the subsequently published
data (de Bernardis et al.\ 2000\markcite{Boom}; 
Hanany et al.\ 2000\markcite{maxima}).

The \boom\ data are well described by a model in which
all cosmological parameters except the geometry are fixed to values measured
by independent means.  Once the position of the first peak is fixed,
no tuning of any of the many other parameters 
is required to explain the low observed amplitude
of the second peak.  This is not surprising; it is
simply what is expected in a purely baryonic universe.

Consideration of a purely baryonic universe is motivated by 
the recent successes (e.g., McGaugh \& de Blok 1998b\markcite{MBb})
of the hypothesized alternative to dark matter
known as MOND (Milgrom 1983\markcite{M83}).
Such a modification to conventional dynamics does appear to be viable.
Taken in sum,
the data suggest a universe in which $\Om = \Ob$ and $\OL \approx 1$.

\acknowledgements I thank Cole Miller for conversations about reionization
and scattering, Rabi Mohapatra for discussions about neutrino masses,
and Glen Starkman for insights into geometry.  I thank Cole
Miller and Bob Sanders for a careful reading of the manuscript, and
Jim Peebles, Greg Aldering, Jerry Sellwood, Eric Gawiser, and Arthur Kosowsky
for their comments.

\clearpage
\begin{deluxetable}{lccccccc}
\tablewidth{0pt}
\tablecaption{Model Parameters and Likelihoods}
\tablehead{\colhead{Model} & \colhead{\Ob} & \colhead{\Oc} & \colhead{\OL} &
\colhead{$\alpha$\tablenotemark{a}} &
\colhead{$\chi_{\nu}^2$} & $P(\chi_{\nu}^2)$}
\startdata
Prior \LCDM\ 1 & 0.010 & 0.200\tablenotemark{b} & 0.790 &
1.00\tablenotemark{c} & 13.34 & $\ll 10^{-3}$ \nl
Prior \LCDM\ 2 & 0.020 & 0.200\tablenotemark{b} & 0.780 &
1.00\tablenotemark{c} & 8.30 & $\ll 10^{-3}$ \nl
Prior \LCDM\ 3 & 0.030 & 0.200\tablenotemark{b} & 0.770 &
1.00\tablenotemark{c} & 4.60 & $\ll 10^{-3}$ \nl
D/H\tablenotemark{d}~ \LCDM\
	& 0.039 & 0.317 & 0.644 & 1.00\tablenotemark{c} & 3.72 &
$\ll 10^{-3}$ \nl
Prior Baryon   1 & 0.010 & 0.000 & 0.990 & 0.55 & 1.90 & 0.05 \nl
Prior Baryon   2 & 0.020 & 0.000 & 0.980 & 0.62 & 0.89 & 0.55 \nl
Prior Baryon   3 & 0.030 & 0.000 & 0.970 & 0.66 & 0.58 & 0.81 \nl
D/H\tablenotemark{d}~ Baryon   & 0.034 & 0.000 & 1.010 & 1.00 & 0.55 & 0.83 \nl
\enddata
\tablenotetext{a}{Geometric scaling factor $\ell \rightarrow \alpha \ell$.}
\tablenotetext{b}{Models with $\Oc = 0.3$ and 0.4 with the same baryon fraction
and $\Ob h^2$ give the same result.}
\tablenotetext{c}{$\alpha \approx 0.93$ gives the best match to the position
	of the first peak.}
\tablenotetext{d}{Adheres to $\Ob h^2 = 0.019$ (Tytler et al.\ 2000).}
\end{deluxetable}

\begin{deluxetable}{lccc}
\tablewidth{0pt}
\tablecaption{Quantitative Measures}
\tablehead{ & \colhead{$\ell_{2}/\ell_1$} &
\colhead{$(C_{\ell,1}/C_{\ell,2})_{abs}$} &
\colhead{$(C_{\ell,1}/C_{\ell,2})_{rel}$}}
\startdata
\LCDM\tablenotemark{a} & $\lesssim 2.4$ & $< 1.9$ & $< 3.6$ \nl
Pure Baryon\tablenotemark{a} & $\gtrsim 2.6$ & $> 2.1$ & $> 5.0$ \nl
Measured\tablenotemark{b} & 2.75 & 2.68 & 7.7 \nl
$2\sigma$ variation\tablenotemark{c} & 2.63 & 2.40 & 5.6 \nl
\enddata
\tablenotetext{a}{Values expected \prior.}
\tablenotetext{b}{Values as measured by \boom\ at each apparent peak
	($\ell_1 = 200$ and $\ell_2 = 550$).}
\tablenotetext{c}{Values measured by making $1 \sigma$ changes to
	each of two strategically chosen data points in the direction
	favoring CDM.}
\end{deluxetable}

\clearpage
\figcaption[Fig1.ps]{The \prior\ predictions of (a) \LCDM\ and (b) purely
baryonic models plotted against the \boom\ data.  The amplitudes of the
models are arbitrary and are scaled to match the amplitude of the first peak.
Solid lines in (a) are 
the \LCDM\ models of McGaugh (1999) with baryon fractions
$f_b = 0.05$, 0.10, 0.15 (\LCDM\ models 1, 2, and 3
of Table 1) in order of decreasing amplitude of the second peak.  
These are illustrative of reasonable \LCDM\ models.
The dotted line shows a reproduction of all the parameters
of ``standard'' \LCDM\ (e.g., Turner 1999).  The low amplitude of the second
peak was unexpected: 
all reasonable variations of the parameters of the \LCDM\ model which were
considered before the \boom\ results predicted a second peak considerably
larger in amplitude than allowed by the data.  In contrast, the data are
consistent with the \prior\ predictions for a purely baryonic universe 
containing no CDM.  The solid lines are identical to the previously published
$\Ob = 0.01$, 0.02, and 0.03
models of McGaugh (1999) with geometry scaled to match the 
position of the first peak (Table 1).  Also shown is a model (dotted line)
with the baryon density given recently by Tytler et al.\ (2000).
The data are consistent with a purely baryonic universe devoid of CDM.}

\clearpage
\begin{figure}
\plotone{Fig1.cps}
\end{figure}


\clearpage
\begin{references}
\reference{BBS} Begeman, K.G., Broeils, A.H., \& Sanders, R.H. 1991,
                \mnras, 249, 523
\reference{Sid} Bludman, S.A. 1998, \apj, 508, 535
\reference{BFPR} Blumenthal, G.R., Faber, S.M., Primack, J.R., \&
	Rees, M.J. 1984, \nat, 311, 517
\reference{CDS} Caldwell, R.R., Dave, R., \& Steinhardt, P.J. 1998,
	\aaps, 261, 303
\reference{Boom} de Bernardis et al.\ 2000, \nat, 404, 955 (\boom)
\reference{BM} de Blok, W.J.G., \& McGaugh, S.S. 1998, \apj, 508, 132
\reference{EB} Efstathiou, G. \& Bond, J.R. 1999, \mnras, 304, 75
\reference{Gus} Evrard, A.E. 1997, \mnras, 292, 289
\reference{FP} Flores, R.A., \& Primack, J.R. 1994, \apj, 427, L1
\reference{maxima} Hanany, S. et al.\ 2000, preprint (astro-ph/0005123)
\reference{HSS} Hu, W., Sugiyama, N., \& Silk, J. 1997, \nat, 386, 37
\reference{lange} Lange, A.E. et al.\ 2000, preprint (astro-ph/0005004)
\reference{mario} Mateo, M.L. 1998, \araa, 36, 435 
\reference{mypred} McGaugh, S.S. 1999, \apj, 523, L99
\reference{MBa} McGaugh, S.S. \& de Blok, W.J.G. 1998a, \apj, 499, 41
\reference{MBb} McGaugh, S.S. \& de Blok, W.J.G. 1998b, \apj, 499, 66
\reference{BTF} McGaugh, S.S., Schombert, J.M., Bothun, G.D.,
	\& de Blok, W.J.G. 2000, \apj, 533, L99
\reference{Ben} Moore, B. 1994, \nat, 370, 629
\reference{MQGSL} Moore, B., Quinn, T., Governato, F., Stadel, J., \& Lake, G.
	\mnras, 310, 1147
\reference{M83} Milgrom, M. 1983, \apj, 270, 371
\reference{7dw} Milgrom, M. 1995, \apj, 455, 439 
\reference{M97} Milgrom, M. 1997, \apj, 478, 7 
\reference{M98} Milgrom, M. 1998, \apj, 496, L89
\reference{M99} Milgrom, M. 1999, Physics Letters A, 253, 273
\reference{Cole} Miller, M.C. 2000, preprint (astro-ph/0003176)
\reference{NS} Navarro, J.F., \& Steinmetz, M. 2000, \apj, 528, 607
\reference{OS} Ostriker, J.P., \& Steinhardt, P.J. 1995, \nat, 377, 600
\reference{PSH} Peebles, P.J.E., Seager, S., \& Hu, W. 2000,
	preprint (astro-ph/0004389)
\reference{SNCP} Perlmutter, S. et al.\ 1999, \apj, 517, 565
\reference{hiz} Riess, A.G. et al.\ 1998, \aj, 116, 1009
\reference{cl1} Sanders, R.H. 1994, \aap, 284, L31 
\reference{S96} Sanders, R.H. 1996, \apj, 473, 117
\reference{S98} Sanders, R.H. 1998, \mnras, 296, 1009
\reference{cl2} Sanders, R.H. 1999, \apj, 512, L23
\reference{S00} Sanders, R.H. 2000, \mnras, 313, 767
\reference{SV} Sanders, R.H. \& Verheijen, M.A.W. 1998, \apj, 503, 97
\reference{Jerry} Sellwood, J.A. 2000, \apj, in press (astro-ph/0004352)
\reference{TZ} Tegmark, M., \& Zaldarriaga, M. 2000, preprint, atro-ph/0004393
\reference{debate} Turner, M.S. 1999, \pasp, 111, 264
\reference{tyt} Tytler, D., O'Meara, J.M., Suzuki, N., \& Lubin, D. 2000,
	\physscr, in press (astro-ph/0001318)
\reference{cmbfast} Seljak, U. \& Zaldarriaga, M. 1996, \apj, 469, 437
\reference{WSP} White, M., Scott, D., \& Pierpaol, E. 2000,
	preprint (astro-ph/0004385)
\end{references}
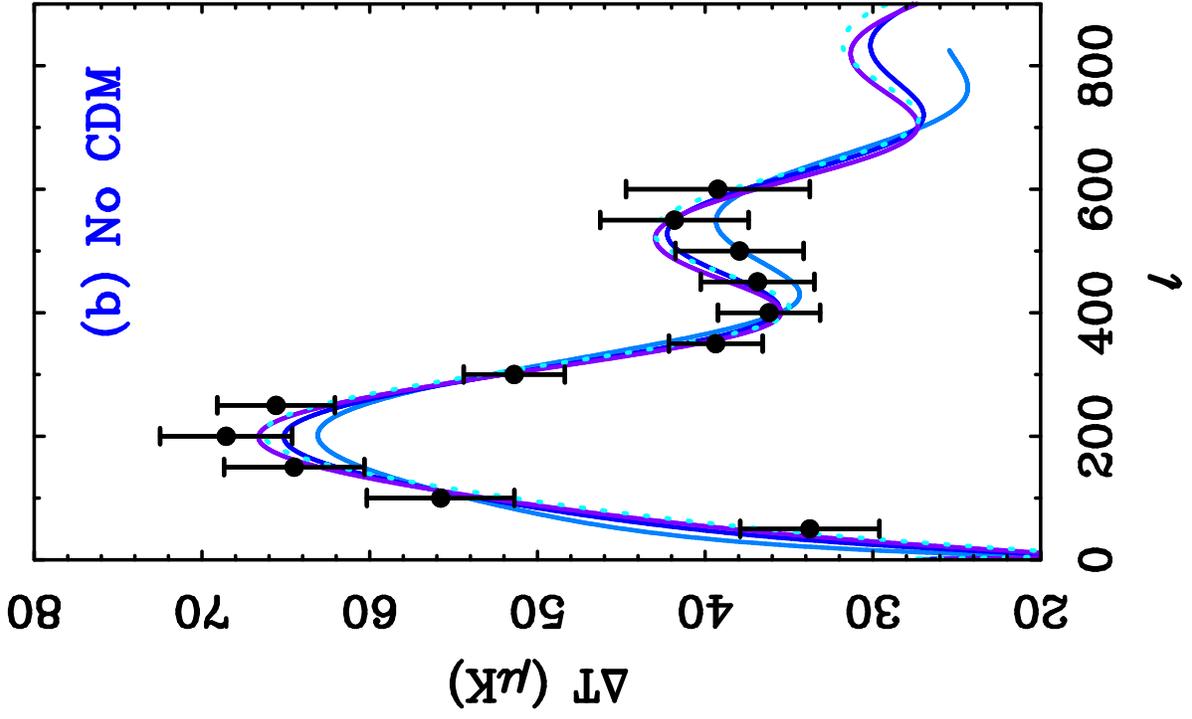
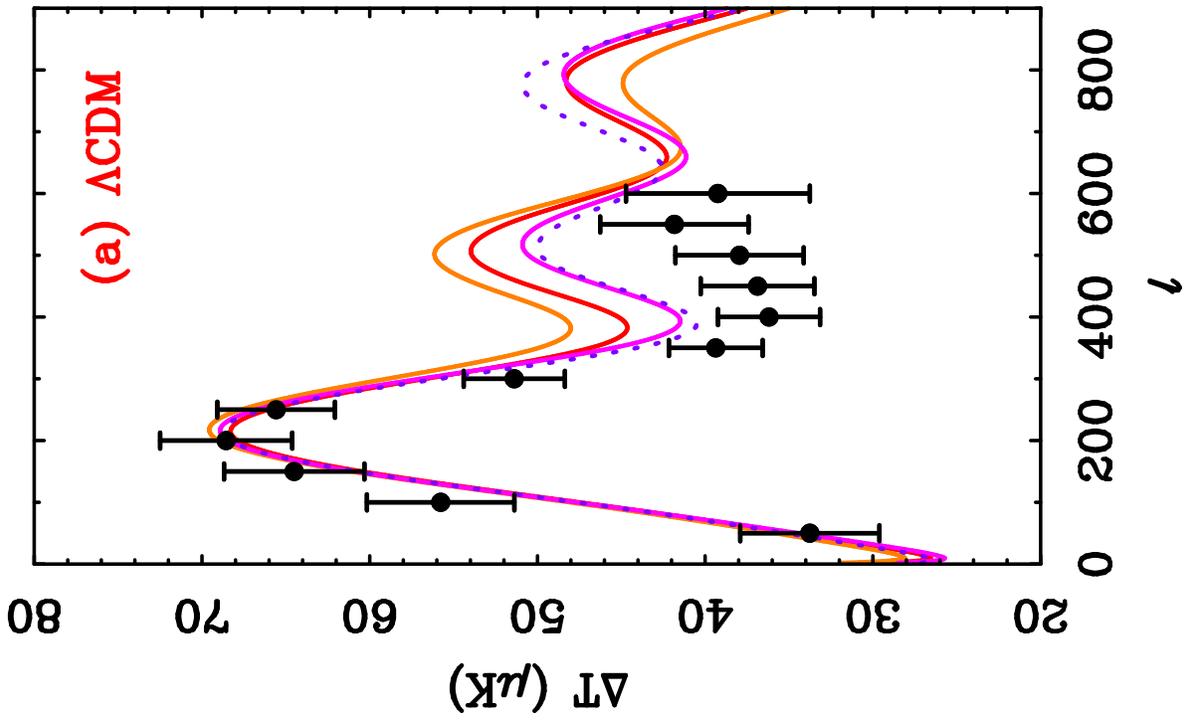
\end{document}